\documentclass[10pt]{article}

\usepackage{
amssymb,epsf,amsmath}

\numberwithin{equation}{section}

\def\de {\partial}
\def\cy {Calabi--Yau}
\def\cldd {Clifford(d,\,d)}

\def\clss {Clifford(6,6)} 

\begin{document}

\setcounter{page}{0}
\begin{titlepage}
\titlepage
\rightline{hep-th/0412086}
\rightline{CPHT-RR 066.1204} 
\vskip 3cm
\centerline{{ \bf \large Modified pure spinors and mirror 
symmetry}}
\vskip 1.5cm
\centerline{\large Pascal Grange and Ruben Minasian}
\begin{center}
\em Centre de physique th{\'e}orique de l'\'Ecole
polytechnique\\Unit{\'e} mixte du CNRS et de l'\'Ecole polytechnique, UMR
7644\\91128 Palaiseau Cedex, France\\
\end{center}
\vskip 1.5cm
\begin{abstract}

It has been argued recently that mirror symmetry exchanges two pure
spinors {\hbox{characterizing}} a generic manifold with
$SU(3)$-structure. We show how pure spinors are modified in the
{\hbox{presence}} of topological D-branes, so that they are still exchanged by
mirror symmetry. This exchange emerges from the fact that the modified
pure spinors come out as moment maps for the symmetries of A and
B-models.  The modification by the gauge field is argued to ensure the
inclusion into the mirror exchange of the A-model non-Lagrangian
branes endowed with a non-flat {\hbox{connection}}. Treating the connection as
a distribution on an ambient six-manifold, assumed to be
$T^3$-fibered, the proposed mirror formula is established by fiberwise
{\hbox{T-duality}}.

\end{abstract}
\vfill
\begin{flushleft}
{\today}\\
\vspace{.5cm}
\end{flushleft}
\end{titlepage}

\newpage



\section{Introduction and conclusions}

Generalized geometry~\cite{hitchin, gualtieri}, which naturally
incorporates the action of a two-form $B$ and has played an important role
in number of recent string-theoretic applications, can be defined in terms
of formal sums of special even or odd forms called pure spinors. For any
manifold of $SU(3)$-structure there is a pair of such objects, the
exponentiated fundamental form $e^{i \omega}$ and the holomorphic
three-form $\Omega$, and the action of mirror symmetry in string
backgrounds (generally speaking with fluxes) can be described as their
exchange. Incorporating D-branes in this picture is of obvious physical
interest. In the large-volume geometrical approach these can be viewed as
gauge bundles supported over some submanifold $Y$ of a manifold $X$. In
the framework of topological strings, D-branes must satisfy some stability
condition in order to correspond to extended objects of untwisted string
theory. Incidentally, the D-brane stability conditions can be compactly
written using the pure spinors.\\

Just as for the pure spinors, the natural distinguishing principle for 
D-branes is whether the volume form of the (sub-)manifold, wrapped by the 
brane,  is of odd  or even rank. For 
the even-dimensional, or B-type, branes one finds on the world-volume the 
deformed Hermitian Yang--Mills equations 
\begin{equation}\label{DHYM} {\rm{Im}}(e^{i\theta} 
e^{i\omega+F})=0.\end{equation}
These were obtained in ~\cite{MMMS} by considering open-string 
effective actions and studying  BPS conditions; similar deformed equations
have emerged in differential-geometric studies of stability 
of vector bundles~\cite{Leung1,Leung2}. They
 correspond to the vanishing of a moment map derived from certain
 symplectic structures on the space of gauge connections. On the other
 hand, world-sheet techniques allow for the identification of
 topological branes with solutions of current-matching
 equations~\cite{OOY}. Stability is demanded in this context by
 matching spectral flows~\cite{Kapustin-Li}, which not only confirms
 the stability equations obtained through space-time techniques for
 holomorphic line bundles, but yields stability equations for
 A-branes, with possible involvement of a non-trivial field strength:
   \begin{equation}\label{stability} {\rm{Im}}(e^{i\theta}\Omega|_Y\wedge F^l)=0,
\end{equation}
 where $Y$ is the world-volume of the A-brane and $l$ is related to
 its transverse complex dimension. The case of non-zero $l$
 corresponds to the coisotropic branes discovered by Kapustin and
 Orlov~\cite{KO}, whereas $l=0$ corresponds to the more familiar
 Lagrangian branes. If one considers an A-brane $Y$ inside a
 Calabi--Yau three-fold $X$, it comes with a bundle whose curvature
 descends to a two-form of the quotient bundle ${\mathcal{F}}Y$
 defined by the following exact sequence:
\begin{equation}\label{sequence} 0\to \mathcal{L}Y\hookrightarrow TY \to {\mathcal{F}}Y \to 0,\end{equation}
where $\mathcal{L}Y=\ker \omega|_Y$. The A-type boundary conditions
mix the metric and the field strength so that $\mathcal{L}Y$ is
annihilated by the field strength. Moreover, $\omega^{-1}F$ defines a
complex structure on the quotient bundle ${\mathcal{F}}Y$. The complex
dimension of ${\mathcal{F}}Y$ is precisely twice the integer $l$ encountered
above.\\

The condition of a Lagrangian cycle $L$ to be special is a BPS condition:  
it consists of the existence of a constant phase $e^{i\theta}$ such that
the ambient holomorphic form, once rotated by this phase, pulls back to a
volume form on the cycle $L$: 
 \begin{equation}\label{SLag}\Omega|_L=e ^{i\theta}{\rm{vol}_L}.\end{equation}
 The
phase $\theta$ encodes the combination of the two supersymmetry generators
that gives rise to the unbroken supersymmetry~\cite{Aspinwall}. Lagrangian
branes satisfying the stability equation, or special Lagrangian branes,
were shown~\cite{lyz} to lead to stable holomorphic line bundles under
Fourier--Mukai transform.\\

 The present paper aims at including the general coisotropic branes in
this mirror-symmetric picture, thus identifying non-Lagrangian
A-branes with non-zero field strength also as mirror counterparts of
B-branes. A way of tying these together with the Lagrangian branes is
to make them fit into the symplectic framework, that was developed on
the A-side by Thomas~\cite{Thomas-thesis,Thomas}. In
\cite{lyz,Thomas} a mirror exhange of moment-map problems for A and B
models was proposed. We find a realization of these moment maps as
modified pure spinors, which allows for extensions of the mirror
picture, by inclusion of not only non-Lagrangian A-branes but also 
 non-vanishing $B$-field.\\

Indeed one would like to show that there exists
a modification of pure spinors by terms involving the field strength
$F$, that yields both the B-type and A-type stability conditions,
(\ref{DHYM}) and (\ref{stability}) respectively, and that the modified
pure spinors are explicitly exchanged under the mirror map.  As they
correspond to maximally isotropic spaces, the pure spinors are
annihilated by half of gamma-matrices of \clss\ algebra. Their
similarity with the Ramond--Ramond fields,  which are also spinors of \clss\ (see
\cite{hassan}) has already been used for the quantitative mirror
symmetry proposals in flux backgrounds \cite{fmt, gmpt}. In the study
of the mirror symmetry with D-branes, we can use the \clss\ picture of
D-brane charges \cite{hm}; note that this can be done with a non-zero
$B$-field (and $H$-flux). From other side, when turning the $H$-flux
off, we should find agreement with previous results concerning mirror symmetry
on \cy\ manifolds with bundles \cite{lyz,Thomas,vafa-m}. In that
sense, turning the flux off provides additional consistency checks for
the generalized mirror conjectures \cite{fmt,vafa,glmw,kstt}. Note that
while for the closed-string case the exchange of pure spinors under
mirror symmetry was verified under the assumption that the ambient manifold is
$T^3$-fibered, it was conjectured and verified by turning Ramond--Ramond fluxes on
and following their transformations, that this assumption may be
eventually dropped. Here we will be verifying the exchange of the
modified pure spinors on the $T^3$-fibered manifolds.\\

The structure of the paper is as follows. In section 2 we consider the 
simple case of D-branes on $T^6$. There are no new results here, but this 
example provides a simple illustration of more general results presented 
in  the bulk of the paper.  There we will work formally, considering 
bundles  supported on a submanifold $Y$ of a Calabi--Yau three-fold $X$. 
In section 3 we extend the treatment of the  special Lagrangian 
A-branes to generic coisotropic ones. In section 4 we present the 
modification of  pure spinors and consider the action of T-duality; we 
also include some  concluding speculations on the underlying closed-open 
 string picture and non-Abelian gauge fields.

\section{From holomorphic line bundles to non-Lagrangian branes}

The ref.~\cite{lyz} establishes the exchange between special Lagrangian
branes and stable B-branes under Fourier--Mukai transform,
starting from special Lagrangian A-branes. In particular, the input data
involve a flat connection, and therefore the curvature on the B-brane 
comes entirely from the shift in the connection induced by the
integrand in the Fourier--Mukai transform. The coisotropic A-branes 
are simply excluded from this picture.
Reversing the process of the derivation and starting from
holomorphic line bundles in the B-model, one may hope for the
appearance of both Lagrangian and non-Lagrangian branes after
performing Fourier--Mukai transform.\\

In the simplest example of a six-torus, it was observed
in~\cite{vanEnckevort} how the curvature of a holomorphic line bundle
influences the dimensionality of the brane obtained by Fourier--Mukai
transform. This computation is instructive and some of its outputs (such as
the dimensionality of the mirror brane) can be extended to other
Calabi--Yau manifolds, assuming the existence of a $T^3$-fibration
along the lines of the Strominger--Yau--Zaslow argument~\cite{syz}. We
therefore review it to set the stage for the fiberwise T-duality that
we will use in section 4, in the framework of generalized complex
geometry (see also ~\cite{BBB,BenBassat}). Let us regard $T^6$ as a 
trivial
fibration $T^3\times T^3$, put coordinates $x^\mu$ on the first
factor, $y^\mu$ on the second one, which we regard as the fiber. We
pick a complex structure by writing $z^\mu=x^\mu+i y^\mu$, and
consider a holomorphic line bundle on $T^6$, or equivalently the
curvature
$$F=F_{\mu\nu}dz^\mu d\bar{z}^\nu,$$ viewed as a distribution
whose support is the B-brane we start with. On a $T^3$-fibered \cy\
manifold, we could follow this path in a local sense, with $x^\mu$ being
 a coordinate in a local chart on the base.  Going to real coordinates makes the
antisymmetric part of the matrix $(F_{\mu\nu})$ appear in the diagonal
blocks, and its symmetric part in the anti-diagonal blocks.
\begin{align}
F &=F_{\mu\nu}(dx^\mu\wedge dx^\nu -dy^\mu\wedge dy^\nu) +
iF_{\mu\nu}( dx^\mu\wedge dy^\nu+ dy^\mu\wedge dx^\nu)\nonumber\\
 &=   F_{\mu\nu}(dx^\mu\wedge dx^\nu -dy^\mu\wedge dy^\nu)+
 iF_{\mu\nu}(dx^\mu\wedge dy^\nu -dy^\mu\wedge dx^\nu) \nonumber \\
 &= F^{(A)}_{\mu\nu}(dx^\mu\wedge dx^\nu -dy^\mu\wedge dy^\nu)+
 iF^{(S)}_{\mu\nu}(dx^\mu\wedge dy^\nu).\nonumber
\end{align}
 As antisymmetric matrices,
 the diagonal blocks have even rank (moreover holomorphicity implies
 that the two ranks are equal). The case where this rank is zero is
 precisely the one considered in~\cite{lyz}, while the case of rank two leads
 to a codimension-one A-brane, as we are going to see.\\

 Let us rewrite the curvature as 
$$F = {\mathcal{A}}_{\mu\nu}(dx^\mu\wedge dx^\nu+dy^\mu\wedge
dy^\nu)+{\mathcal{S}}_{\mu\nu}dx^\mu\wedge dy^\nu,$$ and note at this
point that a mere counting argument would yield the relationship
between this block-structure and the dimension of the T-dual
A-brane. If the antisymmetric block $\mathcal{A}$ was absent, the B-brane
would have a four-dimensional world-volume spanning two directions in
the basis and two in the fiber. This configuration is T-dual to a
three-dimensional world-volume, as we shall see by Fourier
transforming the expression. Turning on $\mathcal{A}$ has the effect
of mixing the $x$ and $y$ coordinates of the support of the curvature,
so that fiberwise T-duality removes one directions and adds two.  Going through
the Fourier--Mukai transformation yields a distribution on the dual
fibration $T^3\times \tilde{T}^3$, with the dual fiber equipped with
coordinates $\tilde{y}$.
\begin{align}
e^{F'}&=\int_{T_{y}^3} e^{dy_\mu d\tilde{y}^\mu} e^F\nonumber\\
& =  e^{{\mathcal{A}}_{\mu\nu}(dx^\mu\wedge dx^\nu)}\int_{T_{y}^3} e^{dy_\mu
d\tilde{y}^\mu} e^{{\mathcal{A}}_{\mu\nu}dy^\mu\wedge
dy^\nu+{\mathcal{S}}_{\mu\nu}dx^\mu\wedge dy^\nu}.\nonumber
\end{align}
If ${\mathcal{A}}$ is zero, then the
curvature $F'$ is also zero, and the (Poincar\'e dual of the) resulting Chern character is
proportional to the distribution
$$\delta(\tilde{y}_\mu-{\mathcal{S}}_{\mu\nu} x^\nu),$$ which situation
corresponds to a Lagrangian A-brane with a flat connection, as anticipated by the counting argument. On the
other hand, if ${\mathcal{A}}$ has rank two, a Gaussian integration can be
performed in terms of an invertible submatrix of ${\mathcal{A}}$, called ${\mathcal{A}}^{-1}$. We are going to
make use of the operation $V^\perp\lrcorner(\cdot )$
defined~\cite{fmt} through
\[
V^\perp \lrcorner(e^{\alpha_1}\ldots e^{\alpha_k}) = \frac1{(3-k)!}
\epsilon^{\alpha_1\ldots \alpha_3} e_{\alpha_{k+1}} \ldots
e_{\alpha_3} \ , k=0\ldots 3\ .
\]
This is essentially a Hodge star on the fiber, except it sends a
$k$-form in the fiber into a $3-k$-vector (a section of
$\Lambda^{3-k}T$).  The Hodge dual of the two-form ${\mathcal{A}}_{\mu\nu} dy^\mu
dy^\nu$ is a one-form on the fiber, to which the modified fiberwise
Hodge star $V^\perp$ associates a vector $(V^\perp\lrcorner {\mathcal{A}})$. In
other words, the support of ${\mathcal{A}}$ in the fiber is orthogonal to the
vector $(V^\perp\lrcorner {\mathcal{A}})$, so that this vector entirely specifies
the support of the block-diagonal part of the curvature, enabling to
perform the integration along the fiber, observing that the result is
weighted by (the Poincar\'e dual of) a one-dimensional delta-function:
$$e^{F'}=\delta\Big{(}{(\tilde{y}_\mu-{\mathcal{S}}_{\mu\nu}
 x^\nu)(V^\perp\lrcorner
 {\mathcal{A}})^\mu}\Big{)}\exp\Big{(}{\mathcal{A}}_{\mu\nu}(dx^\mu\wedge
 dx^\nu)+ \frac{1}{2}{(\mathcal{S}}dx)_\mu( {\mathcal{A}}^
 {-1})^{\mu\nu} ({\mathcal{S}}dx)_\nu\Big{)}.$$ Due to the presence of
 the first factor encoding the support of the Chern character, we will
 be able to treat $F$ as a distribution in the sequel, so that the
 pull-back operation to the world-volume of the brane should be
 automatically built-in. We have therefore exhibited a
 five-dimensional object among the possible Fourier--Mukai
 transformations of holomorphic line bundles on the six-torus. The
 extension of this observation to more general Strominger--Yau--Zaslow
 fibrations and its mirror-symmetric interpretation will be the focus
 of the last section.

\section{Action of Hamiltonian vector fields and non-Lagrangian A-branes}
We start this section by briefly reviewing
the definition of moment maps for the action of a group
$G$ on a symplectic manifold $(M,\omega)$. Let $X$ be a generator of
the corresponding Lie algebra; its action induces a vector field $X^\sharp$. If
this vector field is Hamiltonian, then the corresponding Hamiltonian
function $\mu^X$ is called the moment map associated with the
direction $X$. This is expressed as
$$d\mu^X=\iota_{X^\sharp}\omega.$$ The notion of moment map is a
generalization of the Noether procedure in the context of symplectic
geometry (for an ampler review see~\cite{lectures}).\\

Classical mechanics motivates the terminology, and provides the simplest and best known examples. One can for example compute the moment map
 for the action of $\bf{R}^3$ by translations on $\bf{R}^6$ endowed
 with the usual symplectic form $\omega=dq^i\wedge dp_i$. Consider the
 translation by $\de/\de q^i$. The associated vector field is $\de/\de
 q^i$ and one reads off
$$ \iota{\frac{\de}{\de q^i}}\omega=dp_i,$$ hence the moment map for
translation by $\de/\de q^i$ is the momentum $p_i$, thus recovering the
result of the Noether procedure. The same exercise can be done for the 
action of  $\bf{R}^3$ by rotations, yielding the angular momentum as 
moment map.\\

The notion of moment map, after having emerged in global analysis,  was extended by Atiyah and
Bott~\cite{AB} to gauge theory. In what follows, we are going to consider spaces of
differential forms, in particular gauge connections, as
infinite-dimensional symplectic manifolds (for a review of this framework in the context of topological branes, see~\cite{chapter38}).

\subsection{Symplectic structures on gauge theory in the B-model}
Bundles supported on an even-dimensional submanifold $Y^{(2m)}$ of a
K\"ahler manifold are naturally adapted to symplectic geometry on gauge
theory~\cite{Leung1, Leung2}, in the sense that a natural pairing between
one-forms follows from integrating along $Y$, filling the empty
dimensions with the K\"ahler form:
$$\varpi(a,b):=\int_Y a\wedge b^\ast\wedge \omega^{m-1}.$$ The
symplectic reduction of the space of connections is then the space of
gauge orbits of the Hermitian Yang--Mills connections. This comes from
the moment map associated to the gauge transformations that preserve
the holomorphicity conditions, namely shifts of the connection by
$\de\phi+\bar{\de}\phi$, where $\phi$ is a scalar function on $Y$. We
can check that the moment map in the direction $\phi$ is actually
$F\wedge \omega^{m-1}$, and see that this is only a limiting case of a
more general symplectic structure involving the connection. First let
us see how the following quantity responds to a shift by $h+h^\ast$ in
the connection, where $h$ is a $(1,0)$-form:

\begin{align}
  \mu^\phi(A) &:=\int_Y \phi \, dA\wedge \omega^{m-1}.\nonumber\\  
  d\mu^\phi(A)\,.\,h &=\int_Y \phi \,\frac{d}{dt}{\Big{|}}_{t=0} \Big(
  d(A+th)\wedge \omega ^{m-1}\Big)\nonumber \\
  &=\int_Y \phi \,dh\wedge \omega^{m-1}\nonumber\\
  &= -\int_Y (\de+\bar{\de})\phi\,\wedge h\,\omega^{m-1}\nonumber\\
  &= \iota_{\de\phi+\bar{\de}\phi}\varpi \,.\, h.\nonumber
\end{align}
This means that
$\mu^\phi(A)$ is the component along the generator $\phi$ of the moment
map $\mu(A)$ associated to the gauge transformations around the gauge
configuration $A$. As a whole these components make for
$$\mu(A)=F(A)\wedge \omega^{m-1}.$$ 
The more general case comes from
integrating the wedge-product of two one-forms against a gauge-field
dependent kernel, which reduces to the previous one in the limit of
small field strength.
$$\varpi'(a,b):=\int_Y a\wedge b^\ast \wedge \exp(i\omega+F).$$ In
 order for this to make sense as an Atiyah--Bott pairing, we need the
 form $(\omega+iF)^{m-1}$ to be non-degenerate, which condition is
 implied by the deformed Hermitian Yang--Mills
 equation~(\ref{DHYM}). Going through the same computation as above,
 we see that the moment map is actually the top-form from the power
 expansion of $\exp(i\omega+F)$:
\begin{align}
{\mu^{\phi}}'(A) &:=\int_{Y} \phi\, \exp(i \omega+F(A)).\nonumber\\ 
d{\mu^{\phi}}'(A)\,.\,h &= \int_Y \phi\, \frac{d}{dt}{\Big{|}}_{t=0}\big(\exp(i\omega+(F+tdh)\big)\nonumber\\
 &=\int_Y \phi \, dh \wedge \exp(i\omega+F)\nonumber\\
&=-\int_Y (\de\phi+\bar{\de}\phi)\,h \wedge \exp(i\omega+F)\nonumber\\
&=\iota_{\de\phi+\bar{\de}\phi}\,\varpi'\,.\,h\nonumber\\
\end{align}
These components combine into the following moment map, written as the top-form contribution from the integration kernel
\begin{equation}\label{Bmm}\mu'(A)=\exp( i\omega + F(A)) {\Big{|}}_{\rm{top}},\end{equation} 
whose lowest non-trivial component in the small-$F$ limit is the
moment map $\mu(A)$ obtained from the zero-field strength limit of
$\varpi'$. This fact corresponds to the deformed Hermitian Yang--Mills
equations reducing to the ordinary ones in the same limit. The
symplectic framework therefore parallels the results obtained from
supersymmetry requirements in the B-model.

\subsection{Moment map and non-Lagrangian A-branes}
 Symplectic geometry is involved in the A-model in a way that is more
crucial for the geometry of the branes, but makes the symplectic
structure on gauge theory less transparent (for odd-dimensional
branes, the pairing between one-forms cannot be merely inherited from
the ambient K\"ahler form).  The list of symmetries of an A-brane
includes the Abelian gauge group.  It also contains the action of
Hamiltonian vector fields, because the sequence~(\ref{sequence}) is
preserved by this action, whether or not the brane is Lagrangian. The
annihilation of $\mathcal{L}Y$ by the K\"ahler form $\omega$ is
furthermore entangled with gauge symmetry as we are going to show.\\

  Let $h$ be a smooth real function on $X$. The vector field
$V_h:=\omega ^{-1}dh$ is Hamiltonian in the sense of the symplectic
structure $\omega$, and preserves the kernel of $\omega|_Y$. As
noticed by Kapustin and Orlov~\cite{KO}, it also acts on the
connection by Lie derivation, so that allowed modifications to the
connection are actually more involved than the mere shift by an exact
form:
\begin{equation}\label{deformation}\delta A= \mathcal{L}_{V_h}A= \iota_{V_h}dA+ d \iota_{V_h}A=\iota_{V_h}F+d(
\iota_{V_h}A).\end{equation} The first term vanishes in the case of
 special Lagrangian A-branes, because these are equipped with flat
 connections. In that case we are left with the second term, that
 falls into the class of Abelian gauge transformations. But on more
 general coisotropic branes, the field strength can actually be
 deformed under the action of vector fields. Defining a symplectic
 structure on the tangent space to connections on coisotropic branes,
 thus involving the field strength, can lead to richer deformations
 than in the special Lagrangian case. In particular, the moment map
 can be modified by powers of the field strength.\\
 
 Let us review the derivation by Thomas~\cite{Thomas} of the moment
 map in the special Lagrangian case. Let $L$ bee a special Lagrangian
 submanifold of $X$. There, one considers the variation\footnote{the
 action of the Hamiltonian vector field $V_u$ is the flow $\Phi_t$ it
 generates.} of the integral of the holomorphic form against a scalar function $h$,
  under the action of  a Hamiltonian vector field:

$$V_u:= \omega^{-1}du.$$
\begin{align}
\label{Lagrangian}V_u \int_L h\, \Omega &= \int_L h \frac{d}{dt}{\Big{|}}_{t=0}(\Phi_t
^\ast\Omega)=\int_L h d(\iota_{V_u} \Omega)= \int_L dh\land(
\iota_{V_u} \Omega)\\
&=:\varrho(dh, du ),\nonumber
\end{align}
 where the pairing $\varrho$ is defined by its action on
 one-forms as
$$\varrho(a,b)= \int_L a\wedge (\iota_{\omega^{-1}b} \Omega).$$ When
written in terms of the phase $e^{i\theta}$ of the special Lagrangian
condition(\ref{SLag}), the pairing $\varrho$ may be identified as a
metric on the tangent space to the space of flat connections, since it
is symmetric in its arguments:
 $$\varrho(a,b) =\int_L \cos\theta (a\land\ast b),$$
so that the above functional is the component of the moment map along
the generator $h$. Altogether we obtain for special Lagrangian
A-branes:
$$ \mu={\rm{Im}}\,\Omega|_Y.$$
   
  Going beyond the Lagrangian case, consider for definiteness a
five-dimensional submanifold $Y$ of the \cy\ manifold $X$. The natural
integral on $Y$, in which the genuinely coisotropic deformation of the
connection read from~(\ref{deformation}) can modify the
action~(\ref{Lagrangian}), is the following:
 $$\int_Y h \Big{(}d(\iota_{V_u} \Omega)\land F + \Omega\land d(\delta
A) \Big{)}=\int_Y h \Big{(}d(\iota_{V_u} \Omega)\land F + \Omega\land
d(\iota_{V_u}F) \Big{)}.$$ We see that the piece from ordinary gauge
transformations is annihilated by the derivation, and we are left with
the new deformation of the connection. Integrating by parts we obtain
our candidate for the pairing on the tangent space to the space of
connections on the A-brane:
$$\int_Y h \Big{(}d(\iota_{V_u} \Omega)\land F + \Omega\land
d(\iota_{V_u}F) \Big{)}=\int_Y h\, d\iota_{V_u}\Big{(}\Omega\land F
\Big{)}$$
 $$=\int_Y dh \land (\iota_{V_u}(\Omega\land F)).$$ 
The natural pairing to define at this point indeed  reads
$$ \varrho'(a,b)=\int_Y a\land (\iota_{\omega^{-1}b}( \Omega\land F)).$$
 The proportionality factor between the five-form $\Omega\land F$ and
 the volume form on $Y$ is then the suitable slope for coisotropic
 branes. Its very existence moreover ensures that $\varrho'$ actually
 defines a symplectic structure on the tangent space:
$$ (\Omega\land F)|_Y=: e^{i\theta}{\rm{vol}}_Y.$$ Taking into account
the deformation of non-flat connections thus extends the picture from
special Lagrangian A-branes to coisotropic branes and induces the
following moment map:
\begin{equation}\label{Amm}\mu'(A)= {\rm{Im}}\,(\Omega|_Y\land F).\end{equation} This quantity should be mirror to
 the moment map studied in the B-model, provided stable objects are
 mapped to stable objects by mirror symmetry. The symmetries of
 coisotropic A-branes thus fit into symplectic geometry on gauge
 connections to generate a candidate for a mirror of stable
 holomorphic line bundles, that is not tied by the special Lagrangian
 assumption. This modification of the moment map by gauge fields in
 the A-model provides a differential-geometric explanation of the
 world-sheet result of~\cite{Kapustin-Li}. Let us now turn to an
 explicit verification by T-duality of the exchange
 between~(\ref{Bmm}) and~(\ref{Amm}) under mirror symmetry.

\section{T-duality and exchange of modified pure spinors} 

The above completion of the symplectic picture by non-Lagrangian A-branes
has taught us that for any stable brane supported on a submanifold $Y$ of
a \cy\ manifold $X$, mirror symmetry exchanges differential forms of
maximal degree, where the possible contribution of the gauge fields comes
from the Chern character, namely~(\ref{stability}) in the A-model
and~(\ref{DHYM}) in the B-model. Without the contribution from the Chern
characters, these two quantities reduce to two pure spinors that have been shown in~\cite{fmt} to be
exchanged by mirror symmetry, for any manifolds of $SU(3)$-structure possessing a $T^3$-fibration. 
Combining these two objects together, we arrive at a proposal for the 
mirror exchange between two pure spinors modified by a connection  supported 
on D-brane world-volume. \\

We consider generic six-dimensional manifolds $X$ that admit 
$T^3$-fibrations:  strictly speaking there is no reason to expect that 
backgrounds with fluxes should
admit these (and in some known cases they simply do not), however this
assumption is important calculationally. Note that this assumption could
be relaxed when giving the prescription for the mirror symmetry
with $H$-flux but without branes.\\

As just mentioned, T-duality along the $T^3$-fiber  amounts to
the exchange of the two pure spinors $e^{i\omega}$ and $\Omega$. Since the
exchange still holds in the presence of a $B$-field, incorporating branes into
the picture is not too hard. Indeed, by properly defining the gauge
field $F$ in such a way that it extends over the whole six-manifold $X$
irrespectively of the dimensionality of the brane (working with the
elements of $Q \in K(X)$), we work with the gauge-invariant combination
$B-F$ formally treating it as a $B$-field, while bearing in mind the
T-duality transformation of the gauge fields.

\subsection{T-duality for closed strings}

First, we quickly review the T-duality for the case without gauge
 bundles, following \cite{fmt} (with a slight change of conventions in
 the names of the coordinates, for which we rather follow~\cite{lyz}).
 The six-dimensional manifold will be taken to be a $T^3$-fibration
 over a base $B_3$. Coordinates on the base will be denoted by $(x^1,
 x^2, x^3)$, and those on the fiber by $(y^1, y^2, y^3)$.  All the
 quantities will only depend on the $x$ coordinates. Then
 $i,j,k,\dots$ are used in the three-dimensional $x$ subspace and
 $\alpha,\beta,\gamma,\dots$ are used in the three-dimensional $y$ subspace;
 $a,b,c,\dots$ and $a',b',c',\dots$ are used in the three-dimensional real $x$ and
 $y$ frame spaces. As for the six-dimensional covariant notation, $\mu, \nu,\dots$
 are used in the total six-dimensional space for real coordinates
 ($dx^\mu=(dx^i,dy^\alpha)$), while $m,n,\dots$ are reserved for
 holomorphic/antiholomorphic indices; finally, $A,B,C,\dots$ are
 indices in the total three-complex-dimensional frame space.\\

The metric and $B$-field are given by 

$$ds^2 = g_{ij}\, dx^i dx^j + h_{\alpha\beta}\,e^\alpha e^\beta
      =  G_{\mu\nu} dx^\mu dx^\nu $$
$$ B = \frac{1}{2} B_{ij}\, dx^i \wedge dx^j 
         + B_\alpha \wedge (dy^\alpha + \frac{1}{2}\, \lambda^\alpha)
         + \frac{1}{2} B_{\alpha\beta}\, e^\alpha \wedge e^\beta $$
where $\lambda^\alpha = \lambda_i^\alpha dx^i$, 
$B_\alpha = B_{i\alpha} dx^i$, and we have defined 
$$
e^\alpha \equiv dy^\alpha + \lambda^\alpha\ .
$$ Before passing to pure spinors we need to introduce the two
basic objects, namely a two-form $\omega$ and a three-form $\Omega$
satisfying $\omega\wedge\Omega=0$ and $i\Omega\wedge\bar\Omega
=(2\omega)^3/3!$.  To do this, we start from $(1,0)$ vielbein, which
in turn defines an almost complex structure

\begin{equation}\label{holviel}E^A = i e^a_i dx^i + V^{a'}_\alpha e^\alpha
\end{equation}
where $A=a=a'$ goes from 1 to 3; the corresponding (0,1) vielbein is
$E^{\bar{B}} = \overline{E^B}$.
The holomorphic three-form reads $ \Omega = E^1 \wedge E^2 \wedge E^3
         = \frac{1}{6}\: \epsilon_{ABC}\: E^A \wedge E^B \wedge E^C$
and the fundamental two-form $\omega$  is
 
 $$ \omega = \frac{i}{2}\: \delta_{AB}\: E^A \wedge E^{\bar{B}}
    = -V_{i\alpha}\: dx^i \wedge e^\alpha .$$

T-duality leaves $g_{ij}$ and $B_{ij}$ invariant, while the components of the
metric and $B$-field with legs along the dualized directions transform as 
\begin{equation}
  h_{\alpha\beta} \longleftrightarrow \hat{h}^{\alpha\beta}
\  ; \qquad
  B_{\alpha\beta} \longleftrightarrow \hat{B}^{\alpha\beta}
\ ; \qquad
  B_\alpha \longleftrightarrow \lambda^\alpha 
\ .\label{eq:Td}
\end{equation}

We can now introduce the vielbein $\hat{V}^{a\alpha}$ of the T-dual metric
$\hat{h}^{\alpha\beta}$, that satisfies
$\hat{V}^{a\alpha} \hat{V}^{a\beta} = \hat{h}^{\alpha\beta}$:
$$\hat{V}^{a\alpha}
 = \left( \frac{1}{h+B} \right)^{\alpha\beta} V^a_\beta
 = V^a_\beta \left( \frac{1}{h-B} \right)^{\beta\alpha}$$
where $V^a_\beta$ is naturally the original vielbein. Writing down the inverse
$$\hat{V}^a_\alpha
 \equiv \hat{h}^{\alpha\beta} \hat{V}^{a\beta}
 = (h-B)_{\alpha\beta} V^{a\beta}
 = V^{a\beta} (h+B)_{\beta\alpha}\ .$$
We can complete the T-duality rules by giving the transformations of the
vielbeine:
\begin{equation}
\label{eq:Tviel}
 V^a_\alpha  \longleftrightarrow \hat{V}^{a\alpha}
\  ; \qquad
 V^{a\alpha} \longleftrightarrow \hat{V}^a_\alpha .
\end{equation}

We will mostly work in the case where the $B$-field is purely of
base-fiber
type in frame indices. Transformation (\ref{eq:Td}) shows that this
condition is conserved by T-duality, 
and $\hat{h}^{\alpha\beta}=h^{\alpha\beta}$. Consequently,
$\hat{V}^{a\alpha}=V^{a\alpha}$ and $\hat{V}^a_\alpha=V^a_\alpha$.
T-duality then only amounts to moving fiber indices up and down (still
exchanging $B_\alpha$ and $\lambda^\alpha$ though).\\

First we do the easier case, in which there is neither $B$-field nor $\lambda$ 
twisting of the $T^3$ bundle. The basic idea is that $\Omega$ can be
written in a sense as an exponential of the almost complex structure 
${\omega_\mu}^\nu$ applied to a degenerate three-form $\epsilon_{ijk}dx^i dx^j
dx^k$, that can be thought of as the holomorphic three-form in the large
complex-structure limit. More explicitly, we expand 
$$\Omega=E^1\wedge E^2\wedge E^3$$ 
 using the expression for the holomorphic
vielbein in (\ref{holviel}). We obtain four terms, with $dx^3$, $dx^2
e$, and so on. As in the example of the torus, we use the duality-friendly 
operation $V^\perp\lrcorner(\cdot )$.  Using the fact that 
the lower $e_\alpha$ are indeed vectors $\de_\alpha\equiv \de/\de 
q^\alpha$, we map $\Omega$ into a sum of  $(k,k)$ tensors - objects with 
$k$ indices up and $k$ down, the latter being along the 
fiber. The sum can be
expressed as an exponent of $V_i^\alpha e_\alpha dx^i$, which is the complex
structure.   According to 
(\ref{eq:Tviel}) the action of T-duality  now simply raises and 
lowers the $\alpha$ index: the tangent
bundle (in the fiber direction) of the initial manifold is equal to the
cotangent bundle (again in the fiber direction) of the T-dual manifold.
As a result, the complex structure gets now mapped to $V_{i\alpha}e^\alpha
dx^i$, the fundamental two-form $\omega$, so that 
$$ T(V^\perp \lrcorner \Omega) = \frac i{3!}\: e^{i \omega}\ .$$

To incorporate the case with non-zero $B$-field and corresponding
twisting of the $T^3$-fiber with a connections $\lambda$, it is
convenient to recall that $e^{i\omega}$ and $\Omega$ are \clss\
spinors, and that the forms act on these as combinations of gamma
matrices. In particular, due to purity of $\Omega$, we have the identity $\gamma^m
\Omega=\gamma_{\bar m}\Omega=0$.  Taking $B$ for the time beimg to be only
 of base-fiber type, we note that due to $\gamma^\alpha\Omega = i \gamma^i
V_i^\alpha\Omega$ we have $e^B \Omega = e^{iB_\alpha\wedge
V^\alpha}\Omega$. T-duality then gives
$$\frac i{3!}  T(e^{i \omega}) = V^\perp \lrcorner (e^B\Omega)\ e^{-B_{\alpha}
\lambda^{\alpha}},$$
 $$ T(\Omega)= \frac i{3!} V^\perp \lrcorner (e^B e^{i \omega})\ e^{B_{\alpha}
\lambda^{\alpha}} .$$

This can be presented in a better form as
\begin{equation}
  \label{eq:tdual2}
  T: \,\,\,\,\,\,\, \frac i{3!} e^{i\omega +B} \longrightarrow \int_{T^3} e^{\cal P} e^B \Omega ,
\end{equation}
and the same with $e^{i\omega}$ and $\Omega$ exchanged. We denoted by ${\cal P}= e^{\alpha}
\wedge {\hat e}_{\alpha}$ the connection on the ``twisted" Poincar\'e bundle.
Due to it being inert under T-duality, adding $B_{ij}$ is trivial, so $B$ in
(\ref{eq:tdual2}) has either both legs along the base, or base-fiber components; the case of a 
$B$-field with all legs along $T^3$ leads upon T-duality to non-geometrical
situations and will not be considered here.

\subsection{Open-string T-duality}

Having reviewed the closed-string case, we are ready to turn to the
branes.  Before we do so, we note that for the above construction as
well as for the applications, a crucial use was made of similarity
between pure spinors and Ramond--Ramond forms, both transforming as spinors of
\cldd. In \cite{hm}, the brane charges have also been treated as
elements of \cldd, and our treatment of T-duality will be following
this treatment closely. We will use the coupling of Ramond--Ramond fields to
D-branes to find the suitable modification of pure spinors for the
open-string case. For the time being, we ignore the gravitational
corrections and take $Q= e^F$; we mostly concentrate on the case of
Abelian branes.\\

As mentioned, we treat the gauge curvature $F$ of all branes as a
six-dimensional object. One way of doing so is to put together the
gauge field on the brane and the transverse scalars into a
six-vector. When restricting to the brane, components of $F$ with two
longitudinal indices will naturally be the gauge curvature on the
brane, while components of $F$ with mixed indices stand for the
derivatives of the transverse scalars (covariantized both by the
connection on the normal bundle). Components of $F$ with two
transverse indices would finally be made of transverse scalars only,
and vanish for a single brane with an Abelian connection. In addition
to this, as for the $T^3$-fibered metric, we take the fields to depend
only on (the subset of) the base coordinates $x$. The branes
generically wrap both the base and the fiber directions.\\

Turning on $F$, the first thing we notice is that it passes through
 $V^\perp\lrcorner(\cdot )$ just like the $B$-field, whenever it is of
 base-fiber type.  Indeed, adding $F_{ij}$ components is again
 trivial, while as can be seen from earlier discussion, $F_{\alpha
 \beta}= 0$ in the case of stable Abelian branes, and so we will not
 need to consider these.  Thus, suppressing the $i$ index and ignoring
 momentarily the $F_{ij}$ component, which are inert under T-duality,
 we can write with a slight abuse of notation the gauge curvature as
 $F= (F_{\alpha}, f^{\alpha})$ with a natural constraint $F_{\alpha}
 f^{\alpha}=0$. Once more, we observe that the $i$--$\alpha$ split has
 nothing to do with the D-brane longitudinal--transverse split. In a
 static-gauge picture, $F_{\alpha}$ here would denote the part of the
 field strength with one leg along the base of the $T^3$-fibration and
 with the other along the fiber; $f^{\alpha}$ stands for derivative of
 transverse scalars parameterizing the directions along the $T^3$
 fiber.\\

Under T-duality along all of $T^3$ we simply have $F_{\alpha}
\longleftrightarrow f^{\alpha}$, and $F_{\alpha} V^{\alpha} +
V_{\alpha} f^{\alpha}$ is invariant. Notice that T-duality along $T^3$
sends the connections with even-dimensional support to those with
odd-dimensional ones vice versa (as it should). The outcome of all
this is that we can simply dress both pure spinors in
(\ref{eq:tdual2}) with $Q= e^F$, and have the same exchange of pure
spinors as under the closed-string mirror symmetry:
\begin{equation}
  \label{eq:tdual3}
  T: \,\,\,\,\,\,\,  \frac i{3!} (Q\wedge e^{i\omega} )e^{B} \longrightarrow 
\int_{T^3}  e^{\cal P} e^{B} (Q \wedge \Omega) .
\end{equation}

One may in fact put the field strength $F$ with its dual $\hat F$ into
a single \clss\ object \cite{hm}. We did not need to do this in order
to obtain the explicit transformation above but this is very much the
underlying logic and is important for considering the general
(non-Abelian) case. Note that these considerations naturally lead to
the definition of a generalized complex submanifold as given by a
product of the tangent bundle to the submanifold with the {\it
co-}normal bundle \cite{gualtieri} (see also \cite{kap, zabzine}).
The formula (\ref{eq:tdual3}) is consistent with the conjectures on
mirror symmetry on \cy\ manifolds with bundles \cite{vafa-m}, and may
also be regarded as a general statement on closed-open string
duality. Turning on and off $F$ and $H=dB$ respectively,  one may relate
different geometries with only the $H$-flux or with a single
D-brane. \\

We have been ignoring the gravitational parts of $Q$, but may recall now
that $Q= \sqrt{{\rm \hat A}(X)}e^{iF}$, so that it is tempting to
speculate that the complete $Q$ should be the factor modifying the pure
spinor; moreover one can check that (\ref{eq:tdual3})  is compatible with
the general T-duality transformations on the D-branes charges as elements
of $K$-theory. The exchange of odd-even dimensional world-volumes is
well-understood on $K$-theory level. Here we emphasize once more that the
discussion is at the level of elements of \clss\ and that all the objects are
defined in the bulk. In fact we note that the modified pure spinor 
$\varphi \wedge Q$ formally looks just like the bulk coupling of Ramond--Ramond forms 
 to  brane charges, and in this sense the mirror/T-duality exchange
(\ref{eq:tdual3}) is very natural. \\

As for the stability conditions, we are not pursuing here the question of
possible generalizations of the calibrations and searching for new stable
branes in the flux backgrounds. So taking $H=0$ (while still having
$B \neq 0$ extension of the Fourier--Mukai derivations, which are
typically performed with vanishing $B$-field; however see \cite{Stern} 
for a quantum mechanical model of D-brane and mirror symmetry in the 
presence of a $B$-field), one can check that the 
modified
pure spinors $\varphi \wedge Q$ indeed do give the moment maps~(\ref{Amm}) 
and~(\ref{Bmm}), and the stability equations for both A and B-type 
branes.\\

We conclude by a conjecture about multiple D-brane wrappings, and thus
the inclusion of non-Abelian connections. These may have $F^{\alpha
\beta} \neq 0$ (going to the static gauge again, on the brane
$F^{\alpha \beta} = [\phi^{\alpha},\phi^{\beta}]$, where
$\phi^{\alpha}$ are the scalars along $T^3$). As argued in \cite{hm,
fh-nonab} for couplings to Ramond--Ramond fields $C$ (at least for the case with $B=0$), 
consistency with T-duality would require then replacing the wedge
product $C \wedge Q$ by a Clifford multiplication.  Given that for
pure spinors as well we are dealing with elements of \clss\,  this is a
natural expectation. At the level of world-volume (i.e. stability
equations) this will replace the deformed Hermitian Yang--Mills
equations by their counterpart with Hitchin terms containing
commutators of scalars. On the B side, such equations have indeed been
studied (see \cite{mt}). On the A side, there will be no modifications
containing the scalars for the coisotropic (including special
Lagrangian) branes. However there would seem to appear a new
one-dimensional equation involving $\Omega_{i \alpha \beta}F^{\alpha
\beta}$, which could be non-trivial for manifolds with $b_1(M) \neq 0$, 
and would indeed restore the democracy between all the odd- and
even-dimensional cycles on A and B sides respectively.

\bigskip\bigskip

{\bf Acknowledgments.}  This work was partially supported by
INTAS grant, 03-51-6346, RTN contracts MRTN-CT-2004-005104 and
MRTN-CT-2004-503369, and by a European Union Excellence Grant,
MEXT-CT-2003-509661.  We are grateful to O. Ben-Bassat,
A. Kapustin, P. Mayr, {\hbox{T. Pantev}} and A. Tomasiello for useful
discussions.


\end{document}